\documentclass[aps,10pt,amsmath,amssymb,nofootinbib,a4paper,showpacs]{revtex4}
\usepackage[breaklinks]{hyperref}
\usepackage[latin1]{inputenc}
\usepackage{graphicx}
\usepackage{color}
\usepackage{amsfonts,amsthm}
\usepackage{bm,bbm}
\usepackage[OT2,OT1]{fontenc}
\newcommand\cyr{%
\renewcommand\rmdefault{wncyr}%
\renewcommand\sfdefault{wncyss}%
\renewcommand\encodingdefault{OT2}%
\normalfont
\selectfont}
\DeclareTextFontCommand{\textcyr}{\cyr}

\begin{document}
\title{No alternative to proliferation}
\author{{\bf Daniele Oriti}}
\affiliation{Max Planck Institute for Gravitational Physics (Albert Einstein Institute) \\ Am Muehlenberg 1, D-14476 Potsdam-Golm, Germany, EU \\ daniele.oriti@aei.mpg.de}
\begin{abstract}
We reflect on the nature, role and limits of non-empirical theory assessment in fundamental physics, focusing in particular on quantum gravity. We argue for the usefulness and, to some extent, necessity of non-empirical theory assessment, but also examine critically its dangers.
We conclude that the principle of proliferation of theories is not only at the very root of theory assessment but all the more necessary when experimental tests are scarce, and also that, in the same situation, it represents the only medicine against the degeneration of scientific research programmes.
\end{abstract}
\maketitle
\section{Introduction and preliminary remarks}
In the following, we present some reflections on the role of non-empirical theory assessment in fundamental physics. 
We discuss the role, virtues and dangers of non-empirical theory assessment at a more general level, and then offer our critica appraisal of the specific criteria for carrying out this assessment proposed by Dawid \cite{Dawid-book}. 

\

Before we proceed, it is useful to clarify the perspective and the limitations of our contribution. 

First of all, we will be mostly concerned with issues of methodology of science, and only indirectly touch on the strictly related problems in epistemology. Also, while we will comment on the psychological and sociological aspects of scientific dynamics, arguing that they should be taken into account in discussing its methodology, we will not attempt any detailed analysis of them, relying mostly on insights from the literature and from personal experience. In fact, and this is probably the most important cautionary note, the perspective adopted in this contribution is (inevitably) that of a theoretical physicist working in quantum gravity, not of a professional philosopher of science. 

This should explain, if not justify, some possible naiveté (hopefully, limited) in addressing the complex topic of non-empirical theory assessment, but also somehow re-assure the reader of our genuine interest in the same topic. Indeed, in quantum gravity the issue of non-empirical theory assessment acquires centerstage, due to the disproportion between observational constraints and theoretical constructions. The variety of the theoretical landscape of different quantum gravity approaches forces any quantum gravity theorist to confront the issue of selecting its own scientific path without empirical guidance, resting entirely on (tentative) non-empirical theory assessment. At the same time, like any other working scientist, we are subject to and witnesses of the cognitive and sociological biases which affect the shaping of the community itself as well as the very development of our theories. They are even more prominent in the case of quantum gravity, due again to the lack of experimental constraints, and this may explain our sensitivity to these aspects of non-empirical theory assessment. However, despite the fact that some work on non-empirical theory assessment, and Dawid's in particular \cite{Dawid-book} concerned one specific approach to quantum gravity, i.e. string theory, we will refrain from commenting on string theory itself or other approaches to quantum gravity. We will try instead to keep our opinions on specific quantum gravity approaches to the background, and tackle the issue of non-empirical theory assessment from a purely philosophical point of view\footnote{Having said this, we also believe that some work on the subject, i.e. in particular Dawid's, is based on an overly generous evaluation of string theory and its achievements as a theory of quantum gravity (without diminishing in any way the justified scientific appeal and the many results of the framework), and a consequent dubious application of non-empirical theory assessment criteria to it (god will forgive him for this). To provide a proper justification of this opinion would require a much more careful analysis that would bring us outside the scope of this contribution, so we leave it for a future one.}.  

Our perspective as quantum gravity theorists is also one main reason why we will approach the topic from a somewhat practical perspective. We interpret the question \lq what constitute good non-empirical theory assessment criteria, if any?\rq ~to have both a descriptive and a prescriptive aspect, and to refer to the scientific practice, rather than the realm of pure philosophical reasoning. Indeed, we are interested in which non-empirical assessment practices can be argued to be {\it fruitfully} applied by scientists to achieve progress, thus with an inevitable {\it prescriptive} aspect, implicitly suggesting what scientists \lq could be doing\rq ~to achieve progress. We do not shy away from this aspect of the discussion, exactly because we approach the methodological issue from the scientist's point of view (it is not the philosophers' task to tell scientists what to do, but it is certainly the scientists' duty to try to do it better, also listening carefully to the insights of the philosophers). At the same time, we know very well that any such prescription can only be tentative, and that there is no such thing as {\it the} scientific method, intended as a simple set of rules scientists follow or should follow, or as the defining feature of science  demarcating it sharply from other forms of knowledge. 

\

Since our take on the issue of non-empirical theory assessment is necessarily influenced by our general view on the nature of science, it is probably also useful to spend a few words to make this explicit. We share a view of science as an adaptive system \cite{science-adaptive}, whose progress could be described in terms very close to the evolutionary ones used in biology, in which scientists (and scientific communities) \lq evolve\rq ~their theoretical (conceptual, mathematical) tools to fit in a constantly changing empirical world of phenomena. This analogy could be phrased in a more impersonal manner, speaking instead of theories themselves evolving through selective pressure in the environmental niches defined by empirical facts. In this view, empirical and non-empirical theory assessment constitute such necessary selective pressure, ultimately driving progress. However, this more abstract way of describing the process of scientific evolution is at risk of neglecting the very human (historical and cultural) components of the same evolution, thus missing much of what is really going on\footnote{We take this to be a well-understood lesson of modern philosophy of science, from Kuhn, Lakatos, Feyerabend, Laudan and many others (for a summary, see \cite{history-social}).}. . For this reason, another useful analogy would be in terms of a \lq marketplace of ideas\rq ~about the world, an admittedly very vague image, but one that gives a good intuition of the dynamical, very human, but at the same time very constrained nature of theory development, assessment and spreading. 
Thus, without denying at all or even diminishing the objective nature of scientific understanding \cite{objectivity}\footnote{We believe that the analysis of social aspects enriches our understanding of scientific objectivity, without undermining it, as in the often distorted picture of science of some social constructivists.}, it helps to keep in mind that this understanding is the product of human minds first, and human communities, then. In fact, we feel that another good way to characterize the scientific way of proceeding is as a constant struggle to understand the world while overcoming (or last, keep on a leash) our many cognitive \cite{cognitive} and sociological \cite{sociological, Stanford-1} biases\footnote{By the latter we mean both the subset of the former that have to do with belonging to a group, and the collective effects of group behaviour that may end up being misleading with respect to the goal of a rational understanding of the natural world; they include institutional factors as well as non-institutionalised aspects of group behaviour; see also the more sociological discussion in \cite{lee}.}, which often mislead us. Given the strength and range of such biases, one would even be tempted to {\it define} the scientific method as the set of tools we have developed over time to make it easier to prove ourselves  wrong (as individuals and as communities). Admittedly, a very weak but, we believe, apt version of methodological falsificationism. If theories are to be considered as constantly under siege of both empirical and non-empirical selection pressure, it goes without saying that the {\it temporary} and {\it dynamical} character of scientific explanations should remain centerstage as well, and any contribution to scientific methodology should be founded on it \cite{lakatos}. Finally, it should be clear that the only notion of {\it truth}, in particular the only notion of scientific truth, that this vision of science allows as meaningful is a partial, temporary, and approximate one. We believe that any attempt at theory assessment that is either motivated by the attempt to establish something more than a truth of this type, or implicitly assumes that this is even possible, is actually betraying the very nature of science.  

\

This should suffice as a sketch of our broader point of view on science, which informs our approach to the issue of non-empirical theory assessment. In particular, it should already explain our insistence, in the following, on the importance of {\it theory proliferation}, raised to the level of a methodological principle, the {\it Principle of Proliferation} (PoP). Given the central role it will play in our reflections, let us make more explicit what we mean by this.    

\

The PoP can be loosely phrased as follows: \lq\lq Construct as many alternatives as possible to the current (dominant) theoretical framework, and use the set of all such alternative theories (including the currently favoured one) as the object of (empirical) testing, not any given theory in isolation\rq\rq . The first prescription applies at {\it any} moment in the evolution of a scientific domain, it does not refer only to moments of crisis following the discovery of some observational anomaly, or some conflict between experiments and the currently accepted theory, or other internal difficulty with it. Also, it clearly applies both to the context of theoretical discovery and to the one of theory justification, assuming we want to retain this distinction. The other very important point to stress is that PoP urges to actively {\it construct} and {\it develop} alternatives to any existing theory, no matter how well supported, and does not merely states the need for {\it tolerance} of any such alternative. These alternative theories, in order to be used as prescribed, have to be sufficiently developed and coherent, and should be able, if shown to fit experience (or other assessment criteria) better than the given theory, to replace it. This suggests to devote as much effort as possible to the {\it development} of full-fledged alternative {\it theories}, not just to their invention as plausible alternative hypotheses.

\

The PoP, in various forms, has been discussed and argued for (but also against) by several authors \cite{PoP}, and, most vocally, Feyerabend \cite{Feyerabend-PoP}. As Feyerabend himself has stressed, however, it could actually be traced back to J. S. Mill and understood as part of his philosophical liberalism \cite{Mill}. 

The arguments for PoP in philosophy of science refer for obvious reasons to the empirical assessment of theories. Indeed, it has been argued that the adoption of PoP leads to an increase of testability, for various reasons. To start with, it improves the understanding of each element in the set, by clarifying what would be denied of the \lq established\rq ~theory if one alternative was shown to be correct, and by elucidating the proper meaning of the established theory itself in relation to its alternatives. Second, an alternative theory may suggest a new empirical test of a given theory, that would otherwise be left unimagined (even if it was fully within the conceptual capabilities of the given theory). Moreover, alternative theories provide means to magnify discrepancies of a given theory with observations, for example by asserting such discrepancies as central facts, while they were only negligible possibilities in the given theory. Finally, alternative theories allow to evade the psychological constraints of a given theory, which may even prevent from noticing at all its weaknesses. 

Notice that PoP also discourages the elimination of older \lq disconfirmed\rq ~or simply disfavoured theories, because they can always win back support, and because their content anyway enrich the given \lq established\rq ~theory. And of course, the PoP does not imply that a scientist cannot hold tenaciously to a given theory she finds promising or convincing, and that is furthermore currently corroborated by observations or non-empirical assessment criteria. On the contrary, the \lq principle of tenacity\rq ~(PoT) is the natural balance of the PoP at the level of scientific communities, and maybe even at the level of individual scientists. However, the principle of proliferation goes together necessarily with the principle of tenacity, while the converse is not true: one can hold a principle of tenacity without logically holding also to proliferation; the result however is dogmatism, which hampers progress. Overall, the PoP thus suggests a picture of progress in knowledge as an ever increasing landscape of alternatives, rather then a convergence toward an ideal view, with each alternative forcing the others into greater articulation, and constantly competing, and all together enriching our mental capabilities\footnote{The dynamical combination of these two principles is also at the basis of Feyerabend's general view on methodology and rationality \cite{farrell}}. At the very least, even theories that are long receded into the background provide a \lq measure of development\rq ~for the more contemporary theories. Why this view is in line with the picture of science as an adaptive system (it is a call for protect and enrich \lq biodiversity\rq) or as a marketplace of ideas should be obvious, as it should be obvious why PoP would be regarded as precious by anyone concerned with the cognitive and sociological biases hampering our search for a solid understanding of the world. Indeed, PoP has been advocated by authors holding a naturalistic perspective on science and an \lq evolutionary\rq ~approach to scientific progress and methodology, while being at the same time supported by results in cognitive science and psychology \cite{Churchland-etc}.

The arguments above refer mainly to empirical assessment, but the usefulness of PoP go well beyond it. In fact, the original arguments given by J. S. Mill in support of PoP were not directly focused on empirical tests (so much that some authors have denied that he was referring to science at all \cite{jacobs}).  They are as follows: 1) a rejected theory may still be true (we are not infallible); 2) even when false, it may contain a portion of truth, and only the collision of different views allows the emergence of such truth; 3) a theory that is fully true but not contested will be held as a matter of prejudice or dogma, without a full comprehension of its rational basis; 4) the full meaning of a theory even if true, can only be comprehended by contrast with other views. As argued by Feyerabend and others \cite{Feyerabend-PoP}: 1) and 2) are supported in history of science: a theory may win competition by chance, greater attention devoted to it, etc before competitors have had time to show their strengths or because they are simply temporarily out of steam, only to win it back later; 3) and 4) receive support by what happens when a given theory acquires centerstage: the risk of a decrease in rationality and understanding, since it does not need anymore to produce good arguments or evidence in its support, it may become part of the general education and academic discourse without having been fully understood and without having necessarily solved its basic problems, debates have to be held in its own terms creating additional troubles to opponents or critics, etc. The favoured theory can of course point to its many positive results, but the important point is to evaluate such results, and this should be done in comparison with some (equally) well developed alternative. Proliferation is then not only an epistemological expression of liberalism, but also the necessary ingredient for a rational enquiry into the nature of things.
Therefore, also on the basis of Mill's original, more general arguments, we will argue that PoP is central and absolutely necessary to non-empirical theory assessment.

\section{The need, purpose and risks of non-empirical theory assessment}
Let us now reflect on the issue of non-empirical theory assessment (NETA) at a rather general level, before turning to a criticism of Dawid's analysis in the following section. 

We share the view of several authors that the methodological issue of identifying good (empirical as well as non-empirical) theory assessment criteria is intertwined with the more epistemological issue of scientific underdetermination \cite{underdetermination, underdetermination-2}, in the sense that criteria for theory assessment and selection are limitations to scientific underdetermination, i.e. to the number of allowed alternative theories which are compatible with the same set of observations that corroborate the given one\footnote{This is only loosely speaking, because to be fully compatible with all existing observations may be too strong a requirement, because what counts as a corroborating observation is partly theory-dependent (for a summary of work on this point, see \cite{theory-laden}), and because one never truly eliminates theories that fail to be corroborated by an observation.}. Such limitations to scientific underdetermination are necessary to have any confidence in our theories. This is simply the core of a fallibilist and critical epistemology, since it only amounts to the fact that we can only have (limited, provisional) confidence in theories which we have critically assessed and that survived some form of (natural) selection. However, we do not think that limitations to scientific underdetermination can or should be strong enough to reduce the drift in theory space to a deterministic path, in which each step forward is the only allowed one. In other words, we hold that scientific underdetermination is unavoidable. In line with the broader perspective on science outlined in the previous section, we are actually naturally led to embrace such underdetermination, as something to exploit as much as possible by creativity and ingenuity (which is the essence of the principle of proliferation). After all, it is exactly this underdetermination that makes scientific progress possible in the first place, since it is in the environment created by scientific underdetermination that new theories are found, ready to replace previous ones, and a wider environment increases the chances that we will find in it the conceptual and theoretical tools we need to explain the world better than we currently do. At the same time, as working scientists we are mainly interested in the scientific underdetermination of {\it existing} alternatives to a given theory, including poorly developed ones or less empirically corroborated ones. 
We believe that such alternatives have to be constructed and developed explicitly, at least to some extent, if they have to be taken as reason to be cautious in trusting a given theory, but also if they have to provide the comparative measure of competitive success for the same theory. As a result, we are not convinced that there is any advantage in going beyond local, and transient, underdetermination \cite{Dawid-book, underdetermination}. We do not underestimate how difficult it is, in practice, to construct functioning theories. On the contrary, it is exactly because we are well aware of how difficult it is, that we are less concerned with the existence of merely logically possible alternatives to any given theory, which we basically take for granted. In fact, the PoP stems from the resulting urge to devote as much intellectual energy as possible to this difficult task of constructing theoretical alternatives that can be used for a meaningful assessment of existing theories.
We also stress that scientific underdetermination, the related competition among theories, and the theory assessment on which it is based, should be understood in their temporal, dynamical nature, and embedded in the historical (and very human) scientific development; it is an abstraction to consider them in purely logical and conceptual terms, as if theories were born complete and competing in some fictitious atemporal (and a-social) platonic arena of ideas. It may be a useful abstraction, in some limited case, but it is also a potentially misleading one, especially when it comes to methodological issues. This implies that when we write \lq theories\rq , we really intend {\it research programmes}  \cite{lakatos}, and that the aim of \lq theory assessment\rq ~is to evaluate as objectively as possible their progressing or degenerating status, not any logical, atemporal truth content, or their abstract likelihood of being empirically confirmed. 

\
   
The above applies to both empirical and non-empirical theory assessment. Let us now focus on the latter.  

NETA constitutes indeed a necessary component of the work of scientists, routinely performed, even if sometimes in a rather unconscious manner (like most methodological decisions and most conceptual analysis, especially when observational constraints and mathematical consistency conditions are already stringent). So, it is certainly true that a picture of theory assessment that includes only empirical criteria (what Dawid calls \lq canonical theory assessment\rq) is too limited to really describe how science proceeds \footnote{We do not think this is really a picture that is held by anyone who has reflected on methodological issues, or that has a direct experience with the daily work of scientific communities, though. Therefore, it is at best an abstraction of a more refined understanding, and attacking this view as such is at risk of attacking a straw man.} Therefore, more philosophical work aimed at making more solid its rational basis is certainly needed. It should also not be controversial at all that any non-empirical assessment criterion we may identify is and will always remain subordinate in its scope and weaker in its impact than any form of empirical theory assessment. Subordinate, for the simple reason that scientific theories are attempts at understanding the natural world, and therefore fitting as well as possible the empirical data is part of their defining goal. Weaker, because the empirical constraints are for some rather mysterious reason especially recalcitrant, although not entirely immune, to being subjected to our sociological and cognitive biases; thus it is an historical fact that they have proven much more effective in applying selective pressure against our best theories, than any non-empirical criterion. Because of all this, if it may be true that the role of non-empirical theory assessment is stronger nowadays and in specific scientific sub-fields, we do not think that this amounts to any radical change of paradigm for scientific methodology.

\

Second, we agree with scholars emphasizing the importance of non-empirical assessment, on the fact that this should not be confined to the realm of \lq theory construction\rq ~but it should be properly understood in the context of \lq theory justification\rq , i.e. not as part of the heuristics of discovery, but as a genuine form of theory assessment, alongside (but subordinate to) empirical assessment.  However, one main reason for objecting to this confinement is that the very distinction between context of discovery and context of justification appears very blurred when the dynamical, historical, and very human character of scientific progress is taken into account. This is a standard result of modern philosophy of science \cite{discovery-justification}, and it is also the daily experience of practicing scientists: theories under development are constantly assessed in their results and their promise, and the selective pressure they are subject to is not applied only on the \lq end result\rq ~of theory development, assuming there is one at all, but leads to a constant re-directing and modifying of the same theories, including (albeit more rarely) their core assumptions. This is one more reason to talk about research programmes, rather than theories, as the true object of theory assessment. Given this blurred distinction, on the other hand, proponents of non-empirical theory assessment criteria should be careful in not doing the opposite mistake, and assume that such criteria belong to the context of justification only and are simply applied to the results of theory construction, again as if the assessment was taking place in an abstract world of platonic ideas. Non-empirical assessment criteria {\it guide} theory construction and influence significantly the establishment of any partial consensus around a given theory. This also means that such consensus, whenever present, cannot be taken as the {\it basis} for non-empirical theory assessment, being the {\it result} of it, and being of course influenced by a number of other factors that are not directly classifiable as rational theory assessment (e.g. sociological factors).    

\

Third, the need for a more detailed analysis of non-empirical theory assessment criteria is all the more pressing in research areas like quantum gravity \cite{QGapproaches}, cosmology, and fundamental physics more generally, where observational constraints are fewer and weaker, much work has necessarily to rely on theoretical considerations alone, and where the underdetermination of theory by data is more noticeable and inevitable, due to the difficulties in acquiring such data in the first place. Having said this, one should add some important cautionary remarks, since fundamental physics offers also a vivid example of the true scope and its risks. 

\

There are several puzzling aspects in current fundamental physics and cosmology \cite{puzzling}, due to observations which are anomalous with respect to the accepted theoretical frameworks, dark matter and dark energy being two examples, and we have plenty of data about them. Can they be used then to constrain, e.g., quantum gravity theories? This of course depends on whether they refer to physical phenomena that can be influenced by (if they are not directly originated from) the fundamental quantum properties of spacetime. But this is an issue that it is up to the various quantum gravity theories to decide on, i.e. it is a {\it theoretical} issue, and different quantum gravity formalisms \cite{QGapproaches} will have a different perspective on it (one can exhibit several examples of heuristic arguments or even concrete simplified models that suggest a fundamental quantum gravity explanation of such observational puzzles). It is in fact affected by what a given quantum gravity formalism says about things like separation of scales, the validity of the continuum approximation for spacetime and the regime in which effective field theory applies, to name a few. Moreover, there is by now a very active area of research dealing with \lq quantum gravity phenomenology\rq ~\cite{QGphen}, where the observational consequences of various quantum gravity scenarios. Some of these are effective field theory models incorporating putative quantum gravity effects, usually based on violation or deformations of spacetime symmetries, and/or on the introduction of minimal lengths or other quantum gravity scales; however, in this category one should probably include  also quantum cosmology models and other symmetry reduced or otherwise simplified models of quantum black holes, all producing tentative observational signatures of the underlying quantum gravity structures. One can attribute the failure of such simplified models, and thus of empirical data, to constrain strongly the more fundamental theories to the need to improve our observations, of course, but a big role is also played by the current inability of fundamental frameworks to connect to these effective simplified scenarios (see \cite{testingQGcosmo} for some recent work in this direction, in the context of cosmology). In the end, how much we really need to rely on non-empirical assessment criteria is a theory-dependent fact. If the goal remains to maximise the testability and adherence to the empirical reality of our theories, then also our non-empirical assessment criteria, i.e. the way we select and shape our fundamental theories, indirectly driving their evolution in one direction or another, should be such that they encourage connection with observations. Then, we have to be wary of any criterion that, not only is of a purely non-empirical type (which may be a necessity), but also makes it more difficult to move towards later empirical tests. On the contrary, we should remember that the relation between theories and observations is not a passive one in which theories sit there waiting to be tested by future observations. Rather, theoretical developments themselves, and the non-empirical criteria we adopt to constrain them, may facilitate or delay (or make impossible) such future testing. 
In fact, one could even raise this suggestion to an assessment criterion in itself: the community should favour (devote more resources and efforts to, regard as more promising, trust more as a viable description of nature, etc) a theory or a theoretical development of an existing framework that brings it closer to empirical testing, or that points to new phenomena amenable to empirical test (rather than in the opposite direction). Another criterion in the same spirit would amount to a negative assessment of any theory that fails to connect, even approximately or tentatively, to observations (i.e. fails to make predictions or to suggest testing observations of some of its basic tenets) after many attempts or after a long enough time. Obviously, no such criterion would imply that the theory is not viable, but it would amount to a non-empirical assessment of insufficient promise to be shown as viable in future observations, simply because it will probably fail to connect to them in a solid way. 
From this point of view, the PoP acquires an immediate important status, exactly because it makes it easier to identify possibles ways of testing theories, and encourages their (collective) confrontation with observations. 
The more empirical assessment criteria recede in strength, the more the PoP acquires centerstage and should provide the main basis for theory development and for non-empirical theory assessment. 

\

Facilitating or, on the contrary, making more difficult to move closer to empirical assessment is only one way in which non-empirical theory assessment could be both useful and dangerous. There are more aspects to be considered in this regard. First, good non-empirical assessment criteria should fully embrace fallibilism regarding scientific theories and they should aim at defining some form of \lq non-empirical falsificationism\rq  ~and avoid any attempt at resurrecting at the non-empirical level the dead corpse of verificationism. Now, it is true that, on the one hand, we never truly falsify theories \cite{lakatos} and thus that any such attempt should simply be understood as applying additional (stronger or weaker) selective pressure on proposed or existing theories. It is also true that, on the other hand, any strategy aiming at verifying (falsifying) a theory can be a small step towards falsifying (verifying) it, when it fails. However, the subtle methodological and psychological difference between the two approaches should not be underestimated. In other words, the point of non-empirical theory assessment (as of the empirical one) is not to give us reasons to \lq\lq trust\rq\rq a theory, but to provide weapons to challenge our trust in it. It is far easier to be fossilised in our trust of a theory, especially absent experimental tests, than to challenge it. The danger is not so much the reliance on non-empirical assessment criteria, but the weakening of the skeptical attitude towards our own theories that should accompany them (and the empirical ones). Second, reliable non-empirical theory assessment criteria can only be justified by a careful analysis of the history of science. Detailed case studies as well as broader surveys of scientific practices and many key historical turning points can allow to infer which non-empirical theory assessment criterion has proven useful in the past, on a rather regular basis, and thus it can be hoped to be useful in the present and in the future. The key words, however, are \lq detailed\rq ~and \lq broad\rq. Absent such detailed and broad confrontation of proposed non-empirical theory assessment criteria with the history of science, any attempted inference of the former from the latter is at risk of being convenient cherry-picking. To put it differently, in our analysis of historical scientific developments, just as in the analysis of the current scientific situation in one or another theoretical domain, it is crucial to distinguish rational assessment, i.e. the identification of rational criteria for non-empirical theory assessment that have been used, from post-hoc rationalization of historic developments, that were independent of such supposed criteria. In looking back at history of science, we should pay due respect to it, and avoid the risk of doing the opposite, i.e. understanding history on the basis of present decisions.  

\

If we have to bring together all these main aspects of what should be the basis of good non-empirical theory assessment, it seems to us that  the PoP should be regarded as the foundation of it, just as it has been argued to be crucial for empirical theory assessment, and even more so, due to the peculiarities and dangers of the non-empirical one. This foundational role stems from the fact that non-empirical theory assessment, just like the empirical one, is always comparative, or at least it certainly works best when it is comparative, and it is a drastic simplification to picture the process of empirical testing of a theory as if it occurred in theoretical isolation from competing theories. This is even more true for the confrontation of theoretical frameworks with our favourite non-empirical assessment criteria. Moreover, the very existence of competing theories enhances the strength of our non-empirical assessment criteria. We believe one can see the immediate non-empirical counterpart of the standard arguments in favour of the PoP, recalled above, as a basis for the conviction we have just stated.  

In brief, the PoP is naturally the engine of a methodology of non-empirical theory assessment that is: subordinate to empirical assessment and aimed at promoting it, fuelled by skepticism towards the theories it is applied to and aiming at challenging rather than corroborating them, fully embedded within the historical scientific development and aiming at contributing to its dynamics, rather than fictitiously placed in the abstract realm of logical possibilities. 


\section{Dawid's NETA criteria: a (non-empirical) critical assessment}
The general view on science we have outlined above, and the resulting focus on some specific aspects of non-empirical theory assessment, which we deem crucial, as discussed in the previous section, inform our appraisal of the specific non-empirical theory assessment criteria that have been proposed by R. Dawid \cite{Dawid-book}. 

These are the Unexpected Explanatory Coherence argument  (UEA: \lq\lq If a theory that was developed to explain the phenomenon X, is unexpectedly found to {\it also} explain the independent phenomenon Y, then this theory is more likely to be true.\rq\rq) argument\footnote{We admittedly paraphrase the arguments in a rather rough manner, but not incorrect or unfair. Obviously, \lq true\rq ~in this context can only mean \lq observationally viable\rq.}, the No Alternative argument (NAA: \lq\lq If no alternative to a given theory is found to explain X, then this theory is more likely to be true.\rq\rq) \cite{NAA}, and the Meta-Inductive argument (MIA: \lq\lq If a theory, that is supported by UEA and NAA (or other non-empirical assessment criteria), has been developed within a research program/tradition in which the same non-empirical criteria have proven successful in selecting empirically corroborated theories, then this theory is more likely to be true.\rq\rq). 

While we find Dawid's application of these criteria to string theory extremely generous and not really solid, we share a basic appreciation of all of them, in that they are indeed often employed in the practice of scientists. In particular the UEA and to the NAA are often stated by practicing scientists in support of their favourite theories, while the MIA is only implicitly employed, if at all. 
However, the NAA is rarely stated in such strong form by cautious practicing scientists, but rather in the form of a weaker \lq best among the alternatives\rq ~argument, or in the disappointed \lq only remaining option\rq ~argument. The only community that voices the NAA in such strong form is the string theory community, and, to be fair, only some subsets of it. We will discuss the NAA argument in more details later on, explaining why we fail to see its usefulness, why we see instead its counterproductive aspects, and why we think that the PoP represents instead the only useful core of the chain of reasoning leading to the NAA, and why it should replace it in a sensible non-empirical (part of) scientific methodology.  

\

The UEA refers to a property of many successful theories, and unexpected explanatory power of theories under development is unanimously considered a sign of their promise. We emphasize that, to have any value at all, the UEA applies only {\it after} the successful explanation of phenomenon X is achieved\footnote{For example, one sometimes hears the argument that string theory is a likely explanation of gravity and fundamental interactions, on the basis of its unexpected and beautiful applications in mathematics. This could be an instance of the UEA only after we have agreed that string theory is in fact a good explanation of gravity and other interactions, otherwise the whole argument is a non sequitur.}. We also emphasize that the UEA works in support of the given theory {\it in comparison} with its competitors, not in absolute sense. Indeed, it is a little harder to move from promise to likely observational viability, as Dawid does, and it is not a step that can be taken on the sole basis of the UEA, and in fact it is not a step that scientists take, in general. Dawid explains this point well, mentioning several possible explanations of the unexpected explanatory power of a given theory that do not imply its own observational viability. In particular, we agree with his conclusion that the UEA can used as a non-empirical support of a given theory only in conjunction with another criterion. However, while he identifies this additional criterion to be used together with the UEA in the NAA, we point out that in practice the only way to rule out the alternative explanations for the unexpected explanatory power of a given theory is to proliferate the theoretical approaches to the explanation of the same (theoretical or observational) phenomena that are connected by the theory in question. Thus it is the PoP that truly works in conjunction with the UEA to provide stronger non-empirical support to a given theory. Consider as an example the issue of determining whether the unexpected explanatory power of a given theory, with respect to some phenomenon Y, is really due to some underlying more general principle that is shared by other theories, and so it does not really imply much about the viability of the theory being assessed. It should be obvious that the only way to do so is to imagine, construct and develop alternative theories which incorporate such more general principle, but differ significantly from (or are even incompatible with) the theory being assessed, and see if they have the same explanatory power concerning the phenomenon Y. The extent to which this is possible will give an indication of the strength of the theory in question as an explanation of Y, and it is determined by the non-empirical assessment criteria that will constrain the construction and the development of alternative theories to it (assuming that they feed on the same empirical basis). This point should become even clearer in the following, after our discussion of the NAA, its problems and its relation with the PoP.  

\

The MIA has a different type of problem. In its core, it seems to us that it is basically a formalization of the sensible suggestion to look carefully at the history of science, in particular the one closest to the theory being assessed, to verify if any specific non-empirical theory assessment criterion has proven useful in the past or not. If it has, present theories supported by it are naturally deemed as more promising\footnote{Again, the shift from \lq more promising\rq ~to \lq more likely to be empirically confirmed\rq ~is implicit in the practice of scientists, but subtle  and very hard to justify at the conceptual level.}. The main problem with turning this sensible suggestion into an assessment prescription is that the strength of the prescription is proportional to the depth and accuracy of the historical analysis and to the strength of its conclusions. And the type of historical analysis needed to run the meta-induction, if one wants to do it properly, is extremely difficult and complex. Moreover, as with any induction, it is proportional to the number of case studies that have been analysed. One would need to show convincingly that: a)  the non-empirical theory assessment criteria that one aims to \lq test\rq ~were indeed primarily responsible for the support of a given theory (it is not enough to show that they contributed to that support, among many other factors; and already to show this is non-trivial at all); b) they have an historical track record of success in picking up the right theory, i.e. they did so regularly, in a very high number of cases and, most important, they only rarely failed. It is not enough to show that they did so on a few occasions, and it is the success {\it rate} that matters, not the existence of many instances of success. Obviously, these are indeed very difficult tasks, even for professional historians, let alone philosophers of science, and let alone scientists wishing to pick up useful assessment criteria for their theories (or, more likely, confirm their prejudices on the validity of their own criteria). Another smaller criticism that can be put forward on the MIA concerns the reference to a given \lq research tradition\rq . There are several traditions living side by side in any scientific domain, which have moreover complex inter-relations; and there are many potentially different ways of pushing forward each of them (depending on which aspect of a given tradition plays a central role in the new theory). This means that, even when a specific set of assessment criteria were found to be successful within a given tradition, it may not be obvious at all to what extent the new theory that we wish to assess by the same criteria is truly \lq in the same tradition\rq ~(so that one can run the MIA) any more than alternative theories. Moreover, the risk of anchoring the potential success of a given theory to past successes of a given research tradition has the risk of being a truism, since most theories are further developments of successful paradigms (thus inheriting also most of their methodology, basically by definition (bar the case of true scientific revolutions). If it reduces to this, the MIA is not very useful, because it could apply to any of the features of the given tradition, not specifically to any methodological criterion. The attempt to apply the MIA to the NAA runs into even more serious troubles, in our opinion, coupling problems with historical induction applied to it, to intrinsic problems of the NAA itself.

\

The more serious problems of Dawid's criteria, in fact, relate to the NAA. 

The first problem with the NAA is that it is simply never true that there is only a single theory or model or hypothesis that is able to explain, in principle, a given phenomenon, if one intends, by this, that the whole community of scientists working on explaining that phenomenon was only able to come up with a single such hypothesis or model, after long enough trying. We know this for a fact in the context of quantum gravity research. And we believe that a careful enough analysis of the historical situation of any specific subfield of science, concerned with the explanation of any specific phenomenon, would show that, at the time, there were several competing ideas for such explanation. Some more, some less developed; some with a larger number of followers, others with less consensus around them; some developed within mainstream research directions, some put forward by somewhat lateral sectors of the community; some more in line with accepted paradigms, some based on more heretical (thus risky) suggestions. Recall that the NAA pretends to infer the viability of a theory from the {\it absence} of alternatives, as a matter of logical deduction, and thus one should first of all ascertain with some confidence this absence, if the argument is to be run at all (thus, even assuming that it is valid, which we are going to dispute). We argue that this is never, historically, the case.

\

Does this mean that we contest the fact that sometimes a given community of scholars can be {\it left} with no serious alternative to a given theory, for explaining a given phenomenon? Of course not. It is basically the definition of accepted science that the (large majority of the) community of scholars has ended up agreeing on a given explanation for a phenomenon, at the exclusion of all others. However, this is (and has always been) the result of empirical assessment, performed either as a final way of eliminating (actually, disfavouring) alternatives to a given theory or as a way to further corroborating a theory that had already won a large support, to finally convince the remaining skeptics. In both cases, the empirical assessment was, ultimately, a comparative assessment. Therefore, it was first made possible, and then made stronger, by a previous phase of intense theory proliferation. This is the basis of the PoP. The question is whether there can be an entirely non-empirical counterpart of this situation. Can we have a situation in which \lq there is no alternative\rq ~to a given theory, before any empirical test? 

We can see only two ways in which this can be achieved. None of them, we argue, allow to infer anything, per se, about the observational viability of the theory that is found \lq\lq without alternatives\rq\rq.  

\

One way is by matter of definitions. One can exclude one or more alternative to a given theory because they do not fit a set of previously chosen desiderata or pre-conditions, often amounting to adherence to the established paradigm. There is nothing wrong with this, of course. The tricky part is which specific desiderata should be chosen to define a \lq successful\rq ~explanation of a phenomenon, because this amounts to a specific hypothesis about which aspects of the existing paradigm should be maintained despite the inability to account for some aspect of the world\footnote{Here we take this inability for granted, for the sake of the argument, but one may simply be dissatisfied with the specific way in which the existing theory describes the world, and consider alternatives to it just for this reason.}, and which other aspect should be understood as bearing the responsability for this inability. These desiderata can be of very different types. They can include specific mathematical ingredients that one believes should be part of the sought for theory, e.g. manifest background independence in the case of quantum gravity, or physical requirements we would like to be realised by our theory, e.g. it should describe fundamental interactions, including gravity, in a unified manner as manifestation of a single physical entity, rather than simply in a unified mathematical language. They can also include requirements on the level of mathematical rigor that a compelling physical theory should be up to, or on the number of additional hypotheses it can be allowed to make to fit existing observations (since a newly proposed theory, based on some new hypothesis that is in contrast with a current paradigm, may be at first in contrast with some part of the observational basis of the same paradigm), i.e. on how \lq speculative\rq ~or \lq radical\rq ~it is allowed to be. Also, it is often the case that the very definition of the phenomenon to be explained is itself influenced by a number of assumptions and pre-conceptions, carried over from the existing theories. Because some of these assumptions and pre-conceptions are necessarily to be reconsidered and dropped, when developing a new theoretical framework, our very definition of the phenomenon to be explained by the new framework will often be revised in due process or, at the latest, once a new theoretical framework is adopted to describe it. All this is normal, but it shows that one can always put enough restrictions on the {\it definition} of what we mean by a candidate explanation of a given phenomenon, to be left with a single such explanation. But these are often restrictions to both our imagination and the paths we decide to explore. It is very dangerous to put too many of them, if we are really interested in finding the best possible explanation of a given aspect of the natural world. Conversely, assuming we are in a situation where a single candidate explanation of a given phenomenon is available, it is often enough to lift some of our preconceived assumptions about the phenomenon itself, or how a good explanation of it should look like, to come up with other tentative alternative explanations, to be then explored further. The PoP, but also the usual scientific practice, encourage us to be careful in damping this further exploration by trying to decide in advance how it has to be carried on. If this is the way a lq no alternative situation\rq is realised, we do not have anything that helps establishing the viability of the survivor theory; on the contrary, we have a problem. We should go back a few steps, and let the PoP come to help.

\

The other way in which a \lq non alternative\rq ~situation can be established is by actual comparison of alternatives, followed by the elimination of some of them as not viable or not promising. This can take place in two cases, distinguished by the stage of development of one or more of the alternatives being compared. 

The first case is when one is including in the comparison theories which are severely under-developed, or in the very early stage of their development. These may range from solid and vast theoretical frameworks which have, however, several key open issues in their foundational aspects or in their physical implications\footnote{Personally, we would include in this category most current approaches to quantum gravity.}, to theories which are barely above the status of hypotheses or templates of actual scientific theories or models\footnote{Here we would include all the quantum gravity approaches which do not fit in the previous category.}. It should be clear that the distinction between theories in this category, and their placing across the mentioned range, is a matter of degree, and not entirely objective. 
This is the most common situation in practice, if one includes in the comparison theories in their very early stages of development, or \lq very young research programmes\rq, and even theories barely beyond the level of interesting speculations, also because the former can be obtained from the existing paradigm or accepted partial theories by simply modifying some aspect of them or by pushing some of their features a little further, and the latter can be generated quite easily with a little imagination, and in fact abound in active communities facing some interesting open issue. This is also the case where comparison is the most difficult. How can we compare a large research programme, albeit incomplete, which has most likely already obtained a number of partial results and therefore attracted the interest of a large subset of the community, with a young and small research direction, that is expectedly much less developed and with fewer partial results to its credit? How do we compare a theory that, maybe because it corresponds to an extension of an existing paradigm, has obviously a more solid foundation, with a wilder speculation, presenting many more shaky aspects, but maybe still a promising solution to the problem at hand? how can we make this comparison fair? 
Notice again that we are talking about the comparative {\it dynamics} of theories, not an abstract side-by-side comparison in some logical space. This means, for example, that any partial success of a given theoretical framework can immediately suggest a modification of an alternative to it that allows it to achieve the same success, or the generation of a new hypothesis for the solution of the same problem, which is then added to the landscape to be developed further.
In a case like this, how can we reach a \lq no alternative\rq ~situation? We could decide to set a threshold on the level of development, below which we judge a theory \lq too underdeveloped\rq ~to be compared with the others or to represent a genuine alternative to them. The problem is that any such threshold will be very ambiguous both quantitatively (the \lq level of development\rq ~is not something that can be measured with any precision) and qualitatively (which aspect counts as crucial for considering a theory \lq well-developed\rq? its mathematical foundations? its conceptual clarity? is it worse to lack a clear connection to possible observations or to be unsure about the fundamental definition of the theory itself?). And again, is the {\it level} of development more important than the {\it rate} of development, i.e its progressive as opposed to stagnant situation? or is it more important to estimate the {\it efficiency of its development rate}, i.e. its rate of development corrected by the amount of resources (time, human power, funding) that have been devoted to it? 
In the end, it is not only unfair but simply logically incorrect to infer anything definite about the viability of a theory from the (possibly) true fact of it being the most developed candidate explanation of a given phenomenon. Such a move, far from being logically cogent, amounts to nothing more than noticing how difficult it is to construct a working explanation of a given phenomenon. Exactly because any practicing theoretician knows how difficult it is to come up with a compelling explanation of a phenomenon (especially if we are not talking about effective models but candidate fundamental theories), the correct attitude in face of this difficulty is to refrain from drawing conclusions from the underdeveloped status of an hypothesis or a framework, in absence of observational support of any of the competing hypotheses on the table, and actually devote more work to bring them to the level that makes any comparison meaningful at all\footnote{Let us stress that we are objecting to the inference from \lq well-developed\rq ~or even \lq best developed\rq ~to \lq likely true\rq ~or \lq likely empirically viable\rq, but we do not fail to appreciate the importance of achieving the former status, exactly because it is difficult to achieve it.}.

The second case (which may be obtained after the application of some development threshold in the first case) is when we compare only well-developed theories with each other. Notice that this is maybe fairer, but not much less ambiguous than the first case. In absence of empirical tests, any such comparison involves judgements about which outstanding open issue of each theoretical framework is a serious shortcoming, maybe even preventing it to be considered as a solid candidate for the explanation of a phenomenon, and which can be instead looked at with some indulgence as a temporary shortcoming in an otherwise compelling theory. In addition, some of the open issues of one theoretical framework may relate directly to aspects which are considered {\it defining features} of a successful explanation of the phenomenon we are interested in. For example, we may consider unification of all forces a defining feature of a successful theory of quantum gravity (to be clear, we do not agree with this definition), and thus evaluate negatively a proposed theory of quantum gravity that, while complete in other aspects, has not made such unification a central feature, and thus it is largely underdeveloped in this aspect, even when it incorporates several strategies for achieving such unification. As a consequence, often this second case is at risk of being an instance of the elimination of alternatives by definition that we discussed above. This implies that also this second case provides a very shaky ground for representing a starting point of the NAA.

Finally, let us notice that the more careful formulation of the NAA as: \lq\lq despite a long search for alternatives to the candidate theory A, no alternative has been found\rq\rq , including explicitly a temporal aspect, is, in fact, a restatement of the first case, more often, or of the second. Indeed, it is usually the shorter version of the more precise statement: \lq\lq despite a long search for alternatives to the candidate theory A, no alternative has been found to be equally compelling, in the sense of being equally developed and solid and of fitting all the assumed desiderata\rq\rq.  

\ 

In the end, on both historical grounds (by which we also mean the daily work of theoreticians) and conceptual grounds, we conclude that the very premise of the NAA is hard to justify. We identify the root of the difficulty in the attempt to run non-empirical theory assessment on the logical grounds of abstract theories, as if they came to us fully formed and definite, neglecting the dynamical and permanent-work-in-progress nature of scientific theories. 

Indeed, the main problem with the NAA, even going beyond all the ambiguities involved in setting up its premise, is that it faces the following contradiction: even if the premise (\lq there is currently no alternative to A\rq) was in some sense correct, it is the {\it result} of some provisional non-empirical theory assessment that has been already carried out. Therefore, as far the analysis of non-empirical theory assessment is concerned, the interesting issue is to understand how such assessment had been performed, not the result of it. The theory emerging \lq without alternatives\rq ~as a result of such assessment will be only as compelling, as a scientific theory, as the (non-empirical) criteria used to arrive at it. The NAA therefore ends up portraying itself as a non-empirical assessment criterion when it is at best a re-statement of the {\it result} of some unexplored (and not even formalised) set of non-empirical assessment criteria having been applied at an earlier stage. But then one has to conclude that the NAA is not, and cannot be, a non-empirical assessment criterion itself. Moreover, it is actually at risk of diverting the needed attention to the real (non-empirical) assessment criteria that led to the its premise ("there is no alternative to theory A"), even assuming this premise is an accurate statement of fact (and, we emphasize again, we believe this is basically never the case in such crude terms).

\

Moreover, the premise of the NAA is inevitably a statement of fact about the {\it consensus} reached in a given research community (or a subset of it). As such it risks being nothing more than a sociological observation, which should carry little heuristic power in itself. In particular, and for the reasons presented above, it fails to provide a rational ground for that consensus, which is taken as the starting point of the argument, and has been achieved by other means (which can also be rather contingent \cite{contingency}). On the other hand, the application of the NAA inevitably feeds back on that consensus, reinforcing it. 
This leads to another problem, again resulting from the failure to take properly into account the dynamical and temporal aspects of theory assessment, i.e. how the use of specific non-empirical theory assessment criteria, and in particular the NAA, may feed back on theory development.
We have already discussed, in the previous section, how important it is, in our view, to ensure that our non-empirical assessment criteria are not only well-grounded in the history of science and in the practice of scientists, but also encouraging scientific progress rather than hampering it, favouring the progressive dynamics of research programmes, rather their degeneration \cite{lakatos}.
What do we have instead, in the case of the NAA? We have a consensus that is maybe well-founded for other reasons and that has been achieved by other non-empirical assessment criteria, which we do not explore. We introduce an additional argument that is not much more than a re-statement of the obvious fact that the rational assessment criteria we employ to achieve consensus are designed to improve the likelihood of a given theory to be correct. This has the only effect of reinforcing the existing consensus. In practise, the NAA works as a pure mechanism for strengthening the consensus, rather than a tool for critically examining it. This is exactly the opposite than a non-empirical assessment criterion should do, to facilitate the progressive dynamics of research programmes. Its applications makes it harder, rather than simpler, to identify critical aspects of our favourite theories, or any prejudice that has unnecessarily entered their construction or earlier assessment. And it goes blindly along the flow of any existing sociological or cognitive bias that we may have been prone to in developing our currently favoured theory.  

\

In order to appreciate this aspect in the case of NAA, let us consider one extreme situation, in which however its premise is clearly true. This is the situation in which a community is looking at a new problem, e.g. explaining some new phenomenon, for the first time, and a first candidate explanation for it is proposed. This is usually the simplest extension of the current theoretical framework (that has been by definition mostly successful until then). Now suppose we apply the NAA. This is indeed one (and possibly the only) case in which its premise holds and it is not the result of any previous theory assessment. It would follow from it that the candidate explanation is a bit more likely to be correct (than it would be if alternatives existed). It is obvious that the only thing one is actually assigning some value to, in this case, is the adherence to an established paradigm and a NAA would simply reinforce the competitive advantage that follows from this. This automatic reinforcement is exactly the opposite of what a good non-empirical assessment criterion should do. It should also be clear that the inference from this of any viability of the theory under consideration is questionable to say the least. The NAA appears extremely weak also in its negative consequences. In the same example above, it is obvious that the situation in which a single candidate explanation for the new phenomenon exist is very short-lived, and usually the relevant community comes up with a number of alternative explanation, more or less radically departing from the existing paradigm. The NAA would imply that as soon as such alternatives are produced, the viability of the first proposed theory is ipso facto decreased, not only before any observational test is performed, but also before any other non-empirical assessment criterion is applied to any of them. We do not find this convincing at all. 
The immediate objection to this example, and the doubts it raises, is of course that the NAA is not supposed to be applied at such preliminary stage of theory development, and that its premise asks that no alternative to a given theory is found "after an extensive search". Granted. But we fall back in the previous objections. That no alternative is \lq found\rq ~in the sense that none is proposed and studied is never the case. If one takes this dynamical perspective on theory assessment seriously, one should realize that the main extra element  that \lq time\rq ~brings in the picture is that both the original theory and its proposed alternatives are developed further, criticized and assessed. As a result, some of them may stop progressing fast enough or become less popular. The new ingredients are therefore the additional non-empirical assessment criteria that are employed to determine such progress or lack thereof, and the consequent increase or decrease of popularity. They would be the interesting object of enquiry, as we emphasized above, but the NAA does not add anything to their work. On the contrary, its only effect would be to stifle the development and exploration of alternatives and the reinforcement of any temporary consensus is achieved. It would increase the chances of theoretical impasse rather than growth.

\

In this dynamical perspective on theory development and assessment, what should be the reaction to a situation in which: a problem is left unsolved; a number of theoretical explanations is proposed; all are non-empirically assessed, with one of them emerging as more convincing than the others? The only progressive strategy is to double the efforts to develop such theory to the stage in which it can be empirically tested, to develop its alternatives further to overcome their current difficulties and to offer a stronger challenge to the favourite theory, and to produce even more theoretical alternatives that can offer the same challenge. The strategy should be to create a landscape in which theories push one another forward, force each other to explore hidden assumptions and to solve outstanding issues, and rush towards connecting to the observations. In other words, the most productive strategy for non-empirical theory assessment is the one based on the PoP, even more than in a situation in which observational tests abound. 
If non-empirical assessment criteria end up instead {\it limiting} the range of explored alternatives or the effort in their development, they make it harder to proceed to empirical assessment, i.e. to confrontation with the real world. 

\

A closer look, therefore, shows on the one hand that Dawid's criteria, and in particular the NAA (in the attempt to establish its premise), crucially rest on the PoP at multiple levels, and if they have any strength at all, it is only within a proliferating environment\footnote{This point is probably obvious to Dawid himself, but not to many of those who have been hailing its arguments in the theoretical physics community}. However, we have also argued that they do not have much strength, and that a proper appreciation of the importance of the PoP really undermines them, in particular the NAA. This is because whenever a first stage of proliferation leads to a situation where the NAA could be invoked (merely as a statement of fact, but without much heuristic value), the response should be \lq more proliferation\rq . If this point is missed, as the enthusiasts of Dawid's criteria in the theoretical physics community risk doing, the same criteria may end up instead undermining theory proliferation, non-empirical theory assessment with it, and, in perspective, scientific progress; they may end up encouraging stasis, rather than moving forward towards better future theories.  

\

We feel that it was important to stress the above points, even if some of them may sound obvious to many practicing scientists and to many philosophers of science. Similar points have been raised by other philosophers of science interested in non-empirical theory assessment, most notably P. K. Stanford \cite{underdetermination, stanford-2}. Stanford makes a number of points we agree with, that are directly relevant to the present discussion and undermine the significance of any NAA, but also of the MIA applied to it. Quoting: \lq\lq [...] we have, throughout the history of scientific inquiry and in virtually every scientific field, repeatedly held an epistemic position in which we could conceive of only one or a few theories that were well-confirmed by the available evidence, while the sequent history of inquiry has routinely (if not invariably) revealed further, radically distinct alternatives as well-confirmed by the previously available evidence as those we were inclined to accept on the strength of that evidence.\rq\rq \cite{underdetermination}. In other words, subsequent investigation always produces new evidence for past scientific underdetermination, i.e. new theories that are equally valid on the basis of previous evidence, and always shows how previous non-empirical limitations to scientific underdetermination were too weak. This is an historic fact, and any reflection of non-empirical theory assessment should start from it. To presume the end of this process of continuous discovery of past scientific underdetermination is in many ways tantamount to presume the end of science. No wonder the NAA is proposed in conjunction with a final theory claim\footnote{We do not discuss final theory claims, since it is beyond the scope of this contribution, and too much tied to current issues in fundamental physics. We do not discuss it also because we believe that our criticisms of the NAA make the further implication to a final theory claim less interesting. However, it is important to keep in mind that a NAA leads naturally in the direction of a final theory claim, as Dawid correctly argues.}.
The above consideration from the history of science, which was implicitly underlying our previous discussion, gives a basis for a \lq meta inductive\rq ~support to the central role of the PoP in non-empirical theory assessment, in the same spirit but in exactly the opposite direction of the MIA, when applied in conjunction to the NAA. In the words of Stanford \cite{underdetermination}: \lq\lq [...] the scientific inquiry offers a straightforward inductive rationale for thinking that there typically are alternatives to our best theories equally well-confirmed by the evidence, even when we are unable to conceive of them at the time. \rq\rq . This is raised by Stanford to a \lq New Induction over the History of Science\rq ~(NIoHS) principle in epistemology/methodology of science (that he is also careful to distinguish from Laudan's \lq pessimistic induction\rq  in the context of the debate about scientific realism \cite{laudan-induction}). It leads us to advocate for the following. If the application of Dawid's MIA to concrete cases is difficult because, absent a truly detailed analysis of the scientific context and minute developments of each such concrete case, it is too prone to cherry-picking and to twisting the history to our purposes, historic analysis shows that the logic \lq\lq I cannot conceive any plausible alternative to X, therefore the likelihood of not-X is low\rq\rq is flawed; far too many examples in the history of science can be exhibited when this \lq\lq no possible alternative can exists\rq\rq conviction, however strongly held, was later shown incorrect \cite{underdetermination}.


There are two possible ways out of the above conclusion, and against the NIoHS, both not very convincing to us. One could try to argue that some new theory also enables our imaginative capabilities to exhausts the space of serious possibilities. This is not impossible, but it would have to be based on a serious analysis of our cognitive capabilities, as well as a deep philosophical (epistemological) analysis. As such, it cannot be internal to any specific scientific theory, i.e. one cannot deduce this exhaustion of possible alternatives from internal aspects of a given theory\footnote{This goes directly against the final theory claim based on the structure of string theory \cite{final}}, and certainly the historical fact of absence of alternatives to a given scientific  theory cannot be such argument. Another possible objection would counter that scientific communities are much more powerful in exploring possible alternatives than individual scientists and that modern ones are even more effective than earlier ones; therefore it is not so implausible that they manage to explore all possibilities and that in the end there is really no alternative to a given present theory. Once more, we believe instead that sociological developments have affected the nature and workings of scientific communities in a way that makes the exploration of theoretical alternatives less, not more, encouraged and effective, due to a general push for conservatism. Among the factors producing this result one could cite professionalization, institutionalisation (with rigid machinery of science production, validation and funding), the rise of Big Science (which also favours risk aversion and conservatism, hierarchical organisation, etc) \cite{Stanford-1}. To give just one example, we refer to Cushing's pyramid structure of scientific communities to explain convergence around one approach, which is particularly strong in absence of empirical tests \cite{cushing}.
Therefore one cannot avoid adopting something like the NIoHS and, in conjunction with a cautious MIA, taking seriously its warning against the NAA.

\

In the end, we do not find the NAA, nor the MIA applied to it, very convincing as useful criteria of non-empirical theory assessment. On the contrary, we find it potentially damaging for the healthy progress of scientific communities. For the arguments presented above, we call instead for the PoP to be recognised as the key to non-empirical theory assessment, and the true engine of scientific progress.

\section{Conclusions: no alternative to proliferation}
In this contribution, we have proposed some reflections about non-empirical theory assessment, based on a general view of science as akin to an adaptive system in evolutionary biology, with empirical and non-empirical theory assessment playing the role of the selective pressure that ultimately drives \lq progress\rq ~(in the non-teleological sense of \lq evolution\rq). We also emphasized the other key aspect that our perspective relies on: the very human nature of science, as the product of individuals and communities, which does not contradict but actually improves (when it functions well) its objective content. This emphasis on the human aspects implies also a very careful attention on the social and cognitive biases that are part of theory development and assessment, that assessment criteria should help to fight. This general view suggested also the main ingredients for our analysis of non-empirical theory assessment:  the importance of the dynamical aspects and feedback loops of theory construction and theory assessment, and the central role played by the proliferation of theories in achieving progress, expressed as the Principle of Proliferation (PoP) for scientific methodology. 

\

We tackled first the issue of non-empirical theory assessment in general terms.  We agreed on the link between theory assessment and scientific underdetermination, in which the first aims at reducing the second. However, we recalled that scientific underdetermination is also the humus of scientific progress, which should be based also on tools aiming at expanding the landscape of possible theories that generate the same underdetermination. This is a first general argument for the PoP. 
We argued that non-empirical theory assessment deserve all the possible attention, because it is routinely performed by scientists, so this philosophical analysis can lead to an improvement of both our understanding and practice of science. We have also emphasized that it is important to keep in mind the subordinate and weaker nature of non-empirical theory assessment in comparison with empirical one, in any attempt to identify good non-empirical assessment criteria.  

In order to identify such {\it good} assessment criteria, we have stressed that, since there is an important feedback between any form of theory assessment and theory development, our non-empirical theory assessment criteria should be judged also in terms of the feedback they produce. Also, since the relation between theories and observations is not a static and passive one, theoretical developments and the non-empirical criteria that constrain them may facilitate or delay future testing; therefore we have to be wary of any criterion that makes it more difficult to move towards later empirical tests. Also from this perspective, the PoP should be seen as the key aspect of non-empirical assessment (one of the main reasons why it was first proposed). 

We have also argued that the good non-empirical assessment criteria are those that aim at falsifying theories, rather than confirming them, just like the empirical ones. Thus, non-empirical theory assessment (as the empirical one) should first of all provide weapons to challenge our trust in our best theories, since it is far easier to be fossilised in our trust of a theory, especially absent experimental tests, than to challenge it. Again, the PoP naturally plays a role in keeping alive this moderate skepticism that is at the heart of science.

\

Nest, we turned to a critical analysis of the specific criteria for non-empirical theory assessment proposed by R. Dawid. While we agreed with him that the UEA of a given theory is an important such criterion that is routinely used, we emphasized that its strength lies in a simultaneous extensive use of the PoP to produce alternative theories and in the criteria that are further used to constrain or eliminate them, rather than, as Dawid argues, in the NAA. Thus our only objection to Dawid's UEA, beside an additional emphasis on the PoP, is really a critique of the NAA. 

We argued that the NAA is both poorly supported and dangerous for its feedback on scientific progress. First, it is never really the case that there simply are no alternatives to a given theory, if we intend it as a pure statement of fact and in such absolute terms. Second, any temporary situation of no alternative to a given theory can only be achieved in two ways: by restricting the terms of the problem (either what the phenomenon to be explained really entails or what the explaining theory is supposed to look like), or by the actual application of some (non-empirical) assessment criteria that disfavour the alternatives. These additional criteria may assess the stage of development of the alternative theories as well as their content, and how crucial are the inevitable shortcoming of each of them. This process is necessarily tentative, often ambiguous, and therefore, even when it leads to establishing something like the premise of the NAA, it provides a very shaky ground for drawing any conclusion from it. Most importantly, if the premise of the NAA is the result of applying some non-empirical assessment criteria to a set of alternative theories, it is these criteria that we should identify and analyse. It is their strength that gives any credibility to the resulting theory found to be the only credibile one, and the NAA itself has no additional heuristic value. Third, the NAA rests on the consensus reached on a certain theory in a specific scientific community, but fails to provide any rational for that consensus. Despite this, it becomes inevitably a tool to reinforce that consensus, and risks not being nothing more. In other words, it does exactly the opposite of what a good non-empirical theory assessment criterion should do, to help scientific progress. Once more, the crucial ingredients entering our analysis are the temporal and dynamical aspects of theory assessment and the consideration of how it reacts back on theory development. 

Our brief analysis, in the end, shows on the one hand that also the NAA crucially rests on the PoP at multiple levels, since it has a chance to have any strength at all only within a proliferating environment. However, it also shows that it does not have much strength, and that a proper appreciation of the importance of the PoP really undermines it entirely. We argue instead that whenever a first stage of proliferation leads to a situation where the NAA could be invoked (merely as a statement of fact), the response should be \lq more proliferation\rq . The failure to do so may end up undermining theory development, non-empirical theory assessment with it, and, in perspective, scientific progress.    

The importance of the PoP, the role it plays in non-empirical theory assessment and the consequent undermining of the NAA, find further support in the analysis of Stanford based on the history of science,  that goes exactly in the direction opposite to the NAA, even if it has the same spirit as the MIA by Dawid. We also warned about the difficulties of applying the MIA to any non-empirical theory assessment criteria, stressing how drawing conclusions from the history of science, as the MIA entails, requires a level of detail and broadness of the historic analysis that is hard to achieve. We conclude that, while we do not disagree with the MIA itself, a careful and broad look at the history of science is more likely to undermine the strength of the NAA rather than the opposite.   

\

In the end, our conclusion is simple. In the search for a better understanding of the natural world, experiments and observations are our best ally, proliferation and fair competition between theories are their complement at the non-empirical level. Both feed the natural, reasonable and reasoned healthy skepticism of scientists. So should do any non-empirical theory assessment criteria. Especially when empirical constraints are scarce, the proper reaction should be more proliferation, and the careful use of those non-empirical assessment criteria that facilitate scientific progress by constraining, but also facilitating such proliferation. Scientific progress requires an open and dynamical scientific community, and an open scientific community entails the highest degree of theoretical pluralism.

\section{Acknowledgments} 
We thank R. Dardashti, S, De Haro, V. Lam, C. Rovelli, C. Smeenk, K. Thebault, C. W\"utrich, and especially R. Dawid, for numerous discussions and clarifications on the topics of this contribution.


\end{document}